\documentclass[a4paper, 12pt]{article}
\usepackage[dvips]{graphicx}

\def\ltap{\ \raise.3ex\hbox{$<$\kern-.75em\lower1ex\hbox{$\sim$}}\ }
\def\gtap{\ \raise.3ex\hbox{$>$\kern-.75em\lower1ex\hbox{$\sim$}}\ }

\begin{document}

\title{Quark droplets with chiral symmetry in the Nambu--Jona-Lasinio model}
\author{S.~Yasui$^a$\thanks{yasui@th.phys.titech.ac.jp} and A. Hosaka$^b$\thanks{hosaka@rcnp.osaka-u.ac.jp} \\
\normalsize $^a$Physics Department, Tokyo Institute of Technology,\\
\normalsize Ookayama 2-12-1, Meguro, Tokyo 152-8551, Japan \\
\normalsize $^b$Research Center for Nuclear Physics (RCNP), Osaka University,\\
\normalsize Mihogaoka 10-1, Ibaraki, Osaka 567-0047, Japan}
\maketitle

\begin{abstract}
We discuss the stability of strangelets 
by considering dynamical chiral symmetry breaking.
For quark droplets of finite volume, 
we formulate the Nambu--Jona-Lasinio model 
with a basis set of the quark wave functions in the chiral bag model.
Chiral symmetry breaking for the finite volume bag is discussed 
in a mean field approximation.
Effects of the pion cloud including the chiral Casimir effect 
are investigated.
Physical quantities of the quark droplets 
such as masses and radii are obtained for quark droplets 
of baryon numbers $A \le 5$. 
We also apply our model setting to the nucleon and discuss 
the stability of the quark droplets against nuclei.
\end{abstract}

\section{Introduction}

The study of the strange matter containing 
a large amount of strangeness is one 
of interesting subjects in the hadron and quark physics.
It has been considered that the strange matter can exit to be a stable state 
due to a large number of degrees of freedom of the color and the strangeness 
\cite{Bodmer, Witten, Farhi}.
It was proposed that the strange matter 
could be the true ground state of the matter of the strong interaction. 
A finite volume droplet of the strange matter, which is called 
strangelet, has been also investigated.
In the analysis of the MIT bag model, it was shown that the strangelets were stable particles of large baryon number and small electric charge as compared with normal nuclei \cite{Farhi, Madsen}.

Several scenarios of formation of the strangelets have been considered in various contexts.
The strangelets may be formed in the relativistic heavy ion collisions in the accelerator which can produce the quark-gluon plasma \cite{Rischke, Bielich}.
In the cosmic scale, the strangelets may be formed in the QCD phase transition in the early universe and/or in the collisions of the compact stars containing a rich amount of strange quarks \cite{Weber}.
As a remnant of such processes in the universe, the strangelets may be observed as exotic particles in the cosmic rays.
Indeed, there are several reports on the observation of the exotic particles in the cosmic rays  \cite{Weber, Banerjee, Kasuya, Ichimura, Price, Miyamura, Capdevielle}.
The baryon number of the observed particle is from a few hundreds to thousands, while the electric charge is the order of ten, which are very much different from the ordinary nuclei.

In the early stage of the study of the strangelets, it was assumed that the quark matter was described as almost free Fermi gas confined in a bag, in which chiral symmetry was restored \cite{Farhi, Madsen}.
There, the finite current mass of quarks played only a minor role, because 
the Fermi energy was larger than the current quark masses of the $ud$ and $s$ quarks.
On the other hand, when chiral symmetry is maximally broken, 
the $s$ quarks would not appear in the ground state of the quark matter 
due to the heavy mass in comparison with that of the $ud$ quarks.
By using an effective model of QCD, 
Buballa {\it et al.} showed that the strange matter was not 
the stable state of the quark matter at low density, 
while it could become the ground state at high density,  
more than a few times of the normal nuclear matter 
density \cite{Buballa96, Buballa98, Buballa99}.
At high density, the chiral symmetry of the $s$ quark 
is restored, and the $s$ quark mass becomes 
smaller than the Fermi energy of the $ud$ quarks.
Therefore, the weak decay process is allowed 
for the generation of $s$ quarks in the quark matter.
However, this model analysis was not applied 
to the strangelets of finite volume, 
since the surface effects were not taken into account.

For the discussion of the stability of the strangelets 
with dynamical quarks, the constituent quark model was used in 
Refs.~\cite{Tamagaki1, Tamagaki2, Tamagaki3}.
There, it was concluded that the strangelets were not stable objects due to the large $s$ quark mass.
In order to discuss the dynamical generation of the quark mass, the point-like interaction of the Nambu--Jona-Lasinio (NJL) type interaction was introduced in the quark droplet \cite{Kim, Kiriyama_Hosaka, Yasui, Yasui2}.
There, the quark wave function of the MIT bag model was used as a set of basis functions in the finite volume system. 
In that discussion, it was shown that the chiral symmetry in the $s$ quark sector was restored for the baryon number $A \ltap 10^{3}$.
There, the strangelets were the ground state of the quark droplets.
The masses of the strangelets were also obtained.
It was concluded that the strangelets were not stable objects as compared with the normal nuclei, since the energy per baryon number $E/A$ of the strangelets were found to be larger than the mass of the ground state baryons.

In these previous studies, the pion and kaon cloud around the bag was not taken into account, though it was necessary to conserve chiral symmetry at the bag.
For the quark droplets with relatively small baryon number $A\ltap 10$, 
the MIT bag with a scalar type boundary condition \cite{Bogolioubov, Thomas, Hosaka, Hosaka_Toki_96} does not give a fully chirally symmetric formulations.
In this paper, we discuss effects of the pion cloud on the chiral symmetry breaking in the quark droplets for small baryon number by using the chiral bag model.

The contents of this paper are as follows.
In Section 2, we introduce the NJL lagrangian with a basis set of the chiral bag model as an effective model of the quark droplet.
In this paper we call this hybrid model as the NJL chiral bag model.
In Section 3, we show the numerical results and discuss the stability of quark droplets.
Section 4 is devoted to the conclusion.

\section{Model}

\subsection{NJL chiral bag model}

We use the NJL type interaction as an effective theory for dynamical chiral symmetry breaking \cite{Nambu}.
For the construction of the quark wave function 
in a finite volume, we consider a bag model for the quark droplets.
Here, we consider the chiral bag model to conserve chirality at the bag surface \cite{Thomas, Hosaka, Chodos, Inoue}.
Then the quarks inside the chiral bag are interacting through the four-point interaction of the NJL model, and the meson cloud exists outside the bag.
The idea of a chiral bag model with the NJL interaction was first presented by T. Kunihiro in 1984 \cite{Kunihiro1, Kunihiro2}\footnote{The authors thank to Prof.~T.~Kunihiro for comments.}.
In this paper, we call this hybrid model as the NJL chiral bag model.
The model lagrangian is given by
\begin{eqnarray}
 {\cal L} &=& \left[
                         \bar{\psi}( i \partial\hspace{-0.2cm}/ - m_{0})\psi
					  + \frac{G}{2}\sum_{a=0}^{8} 
					              \left\{   \left( \bar{\psi}\lambda_{a}\psi \right)^{2}
								              + \left( \bar{\psi}i\gamma_{5}\lambda_{a}\psi \right)^{2}
							      \right\}
                   \right]\theta(R-r)
 \label{eq : NJL_bag} \\ \nonumber
           &&-\frac{1}{2}\bar{\psi} U^{\gamma_{5}} \psi \delta(r-R)
\\ \nonumber
          &&+ \left[ -\frac{f_{\pi}^{2}}{4} \mbox{tr} \left[ \partial_{\mu} U U^{\dag} \partial^{\mu} U U^{\dag} \right] 
          + \frac{1}{32e^{2}} \mbox{tr} \left[ \partial_{\mu} U U^{\dag}, \partial_{\nu} U U^{\dag} \right]^{2}
          + {\cal L}_{mass} \right] \theta(r-R),
\end{eqnarray}
where $\psi=(u,d,s)^{t}$ is the quark field, and the current mass matrix $ m_{0} = \mbox{diag}(m_{0u},m_{0d},m_{0s}) $.
The second term in the first bracket is the NJL type interaction term invariant under $U(N_{f})_{L} \times U(N_{f})_{R}$ with $N_{f}=3$, in which $\lambda_{a}$ ($a=0, \cdots, 8$) is the Gell-Mann matrices normalized by $\mbox{tr}\lambda_{a}\lambda_{b}=2\delta_{ab}$.
We do not consider the 't Hooft term for explicit $U(1)_{A}$ breaking.
The step functions are multiplied in order to distinguish the quark phase inside the bag and the meson phase outside the bag.
Here, $r$ is a distance from the center of the bag.
We assume that the strangelet has a spherical shape with a bag radius $R$.
The second term with the delta function realizes the chiral boundary condition for vector and axial vector current flows at the bag surface with keeping chiral symmetry, in which we define
\begin{eqnarray}
 U^{\gamma_{5}} = \frac{1+\gamma_{5}}{2}U + \frac{1-\gamma_{5}}{2}U^{\dag}.
\end{eqnarray}
The third term is the meson lagrangian outside the bag, where the meson field $\vec{\phi}$ is
\begin{eqnarray}
  U &=& e^{i\vec{\lambda}\cdot\vec{\phi}},
\end{eqnarray}
with
\begin{eqnarray}
   \frac{1}{\sqrt{2}}  \vec{\lambda}\cdot\vec{\phi}  =
\left(
\begin{array}{ccc}
 \frac{1}{\sqrt{2}}\pi^{0} + \frac{1}{\sqrt{2}}\eta_{8} & \pi^{+}  &  K^{+}  \\
 \pi^{-} &  -\frac{1}{\sqrt{2}}\pi^{0} + \frac{1}{\sqrt{2}}\eta_{8}  &  K^{0} \\
 K^{-} & \overline{K}^{0}  &  - \frac{2}{\sqrt{6}}\eta_{8} 
\end{array}
\right).
\end{eqnarray}
The meson fields are normalized by the pion decay constant $f_{\pi} = 0.093$ GeV.
We introduce the Skyrme term to obtain stable solutions~\cite{Skyrme, Adkins}, 
where 
the coefficient is set $e=6.17$ so as to 
reproduce $E_{N+\Delta}=1.1$ GeV, 
the average mass of $N$ and $\Delta$.
The meson mass term is given by
\begin{eqnarray}
 {\cal L}_{mass} = \mbox{tr} M(U+U^{\dag}-2I),
\end{eqnarray}
where $M \sim m$.
In the following discussion, however, we turn off the pion mass term which does not change the result qualitatively.

In the meson sector, we consider only the $\pi$ meson by neglecting $K$ and $\eta$ mesons, because they are heavier than the $\pi$ meson. Therefore, the meson field is written as
\begin{eqnarray}
  U =
\left(
\begin{array}{cc}
 u_{\pi} & 0 \\
 0 &  1    
\end{array}
\right),
\end{eqnarray}
where we define $u_{\pi} =e^{i\vec{\tau}\cdot\vec{\pi}}$.

In the quark sector, chiral symmetry breaking is caused by the non-perturbative effect in the NJL type interaction, which gives a non-zero expectation value of the quark scalar condensate $\langle \bar{q}q\rangle$.
In the NJL interaction, we consider the mean field approximation $(\bar{q}q)^{2} \rightarrow 2\bar{q}q \langle \bar{q}q\rangle - \langle \bar{q}q \rangle^{2}$ for $q=u$, $d$ and $s$, respectively.
We define the dynamical quark mass
\begin{eqnarray}
  m_{q} = m_{0q} -2G\langle \bar{q}q\rangle
\end{eqnarray}
for each flavor.
Note that $m_{0q}=0$ for $q=u$, $d$, but $m_{0s}\neq 0$ for $q=s$.
Then, the lagrangian in the quark sector in Eq.~(\ref{eq : NJL_bag}) is written as the sum of the $ud$ and $s$ quark sectors
\begin{eqnarray}
  {\cal L}_{q} = {\cal L}_{ud} + {\cal L}_{s},
\end{eqnarray}
where ${\cal L}_{ud}$ is the lagrangian in the $ud$ quark sector
\begin{eqnarray}
 {\cal L}_{ud} &=&
                      \left[
                         \bar{h}( i \partial\hspace{-0.2cm}/ - m_{h})h
					  - \frac{(m_{h}-m_{0h})^{2}}{4G}
                   \right]\theta(R-r)
                 -\frac{1}{2}\bar{h} u_{\pi}^{\gamma_{5}} h \delta(r-R)
\label{eq : NJL_bag_mf}
\end{eqnarray}
and ${\cal L}_{s}$ is the lagrangian in the $s$ quark sector
\begin{eqnarray}
 {\cal L}_{s} &=& \left[
                         \bar{s}( i \partial\hspace{-0.2cm}/ - m_{s})s
					  - \frac{(m_{s}-m_{0s})^{2}}{4G}
                   \right]\theta(R-r) - \frac{1}{2}\bar{s} s \delta(r-R).
 \label{eq : NJL_bag_s}
\end{eqnarray}
In Eq.~(\ref{eq : NJL_bag_mf}), 
we define $u_{\pi}^{\gamma_{5}} = e^{i\vec{\tau}\cdot\vec{\pi}\gamma_{5}}$.  
The $ud$ quarks interact with the pion of the hedgehog solution, 
while the $s$ quarks do not.
We define the hedgehog wave function $h$ in the $ud$ sector as discussed below.
We set $m_{h}=m_{u}=m_{d}$ and $m_{0h}=0$.

\subsection{Hedgehog solution}
We assume that the meson field and the interaction are strong, and therefore
the non-linear effects should be included to all orders.
In order to consider the non-linear effect, we assume 
the hedgehog ansatz for the $\pi$ meson sector.
The hedgehog ansatz conserves the grand spin $\vec{K} = \vec{J} + \vec{I}$ with total angular momentum $\vec{J}$ and isospin $\vec{I}$.
The hedgehog pion field for $r>R$ is then written as
\begin{eqnarray}
  \vec{\pi} = F(r) \vec{n},
\end{eqnarray}
with the chiral angle $F(r)$ as a function of $r$ and a unit radial vector $\vec{n}$ in the 
coordinate space.
In the hedgehog ansatz, the unit radial vector is identified with that in the isospin space.
Then, we obtain the energy of the pion field
\begin{eqnarray}
   E_{\pi}(R,F) =  \int_{r>R} d^{3}r
                     \left[
                            \frac{f_{\pi}^{2}}{2} \left( F'^{2} + 2\frac{\sin^{2}F}{r^{2}} \right)              
                         + \frac{1}{2e^{2}} \frac{\sin^{2}F}{r^{2}} \left( \frac{\sin^{2}F}{r^{2}} +2F'^{2} \right)
                    \right].
\end{eqnarray}
Here, $F$ in the left hand side is the chiral angle at the bag surface.
The profile function $F(r)$ is obtained by solving the equation of motion
\begin{eqnarray}
  F'' + \frac{2}{r}F' - \frac{\sin 2F}{r^{2}}
  + \frac{2}{e^{2}f_{\pi}^{2}r^{2}} \left( \sin^{2}F F'' + \frac{\sin2F}{2}F'^{2} - \frac{\sin^{2}F\sin2F}{2r^{2}} \right) = 0,
\end{eqnarray}
which is given by taking a variation of the energy $E_{\pi}$ with respect to $F(r)$.
The asymptotic form of this solution in the limit of large $r$ is $F(r) \sim C/r^{2}$ with a constant $C$.

In the chiral bag, the pion carries a fractional baryon number
\begin{eqnarray}
  B_{\pi} = -\frac{1}{24\pi^{2}} \int_{R}^{\infty} d^{3}x \epsilon_{ijk} 
          \mbox{tr} \left( \partial_{i}U  U^{\dag} \partial_{j}U  U^{\dag} \partial_{h}U  U^{\dag}\right),
\end{eqnarray}
which is characterized by a topological property.
By substituting the hedgehog solution, the fractional baryon number is given by
\begin{eqnarray}
  B_{\pi} = -\frac{1}{\pi} \left[ F -  \frac{1}{2} \sin 2F \right],
  \label{eq : meson_baryon_number}
\end{eqnarray}
where $F$ is a chiral angle at the bag surface \cite{Adkins}.

In the $ud$ quark sector, we consider a basis set for the quark wave function by using a hedgehog solution $h$.
We construct the quark wave function $\psi^{(\kappa)}$ for 
natural ($\kappa=+$) and unnatural ($\kappa=-$) 
parity assignment for the grand spin $K$, respectively;
\begin{eqnarray}
  \psi^{(+)} = 
\left(
\begin{array}{c}
 a_{0} j_{K}(pr) | 0 \rangle + a_{1} j_{K}(pr) | 1 \rangle \\
 a_{2} j_{K+1}(pr) | 2 \rangle + a_{3} j_{K-1}(pr) | 3 \rangle
\end{array}
\right),
\end{eqnarray}
and
\begin{eqnarray}
  \psi^{(-)} = 
\left(
\begin{array}{c}
 b_{2} j_{K+1}(pr) | 2 \rangle + b_{3} j_{K-1}(pr) | 3 \rangle \\
 b_{0} j_{K}(pr) | 0 \rangle + b_{1} j_{K}(pr) | 1 \rangle
\end{array}
\right).
\end{eqnarray}
$j_{K}(pr)$ is a spherical Bessel function with the $ud$ quark momentum $p$.
Here the basis states are given by
\begin{eqnarray}
  | 0 \rangle &=& Y_{KM}(\theta, \phi) \chi_{0}^{0},
\\ \nonumber
  | 1 \rangle &=& \sum_{\mu=-1,0, 1} ( K M-\mu 1 \mu | K M ) Y_{K M-\mu}(\theta, \phi) \chi_{\mu}^{1},
\\ \nonumber
  | 2 \rangle &=& \sum_{\mu=-1,0, 1} ( K+1 M-\mu 1 \mu | K M ) Y_{K+1 M-\mu}(\theta, \phi) \chi_{\mu}^{1},
\\ \nonumber
  | 3 \rangle &=& \sum_{\mu=-1,0, 1} ( K-1 M-\mu 1 \mu | K M ) Y_{K-1 M-\mu}(\theta, \phi) \chi_{\mu}^{1},
\end{eqnarray}
where $Y_{K \, M}(\theta, \phi)$ is a spherical harmonics with spherical coordinate $(\theta, \phi)$.
$\chi_{\mu}^{G}$ are eigenstates of the sum of spin and isospin $\vec{G} = \vec{S} + \vec{I} = \vec{\sigma}/2+\vec{\tau}/2$,
\begin{eqnarray}
 && \chi_{0}^{0} = \frac{1}{\sqrt{2}}( | \uparrow \rangle  | d \rangle - | \downarrow \rangle  | u \rangle ),
\\ \nonumber
 && \chi_{1}^{1} = | \uparrow \rangle  | u \rangle,
\\ \nonumber
&&  \chi_{0}^{1} = \frac{1}{\sqrt{2}}( | \uparrow \rangle  | d \rangle + | \downarrow \rangle  | u \rangle ),
\\ \nonumber
&&  \chi_{-1}^{1} = | \downarrow \rangle  | d \rangle.
\end{eqnarray}
The sign of naturalness and unnaturalness corresponds to the parity $P=(-)^{K+\kappa}$.
The coefficients $a_{i}$ and $b_{i}$ ($i=0,\cdots,3$) are determined by satisfying the equation of motion of the $ud$ quark.

The eigenenergies of the hedgehog $ud$ quarks are given by the boundary condition at the bag surface $r=R$,
\begin{eqnarray}
  i\vec{n}\cdot\vec{\gamma} \psi^{(\kappa)} = - e^{i\vec{\tau}\cdot\vec{n} F(R)\gamma_{5}} \psi^{(\kappa)},
  \label{eq : boundary_vector_1}
\end{eqnarray}
which represents the continuity of the vector current flux.
Then, we obtain the eigenvalue equation which is explicitly shown by
\begin{eqnarray}
\label{eq : boundary_vector_2}
   && \cos F(R) \left[ j_{K}(pR)^{2} - 
   \left(\frac{E-\kappa m_h}{p}\right)^{2} j_{K+1}(pR) j_{K-1}(pR) \right]
 \\ \nonumber
 &&- \kappa \frac{E-\kappa m_h}{p} j_{K}(pR) 
 \left( j_{K+1}(pR) - j_{K-1}(pR) \right)
\\ \nonumber
 &&+ \frac{E-\kappa m_h}{p} \frac{\sin F(R)}{2K+1} j_{K}(pR) 
 \left( j_{K+1}(pR) + j_{K-1}(pR) \right) = 0,
\end{eqnarray}
with the quark energy $E = \sqrt{p^{2}+m_{h}^{2}}$.
The equation for $K=0$ is obtained by setting $j_{-1}(pR)=0$.
We note that there is a symmetry between energy levels 
in the positive and negative energy sides, 
\begin{eqnarray}
   E_{K^{P}}(F) = -E_{K^{-P}}(-F).
\end{eqnarray}
This relation stemmed from the invariance of the lagrangian (\ref{eq : NJL_bag}) under the transformation $U \rightarrow U^{\ast}$ or $F \rightarrow -F$.

In the chiral bag model, the $\pi$ meson cloud at the bag surface induces the vacuum polarization in the quark sector, and consequently the change of the baryon number carried by the vacuum quarks.
In the conventional chiral bag model with massless quark, it has been known that the fractional baryon number of the vacuum quarks is canceled by that of the pion cloud \cite{Goldstone, Mulders}.
We have checked that this is also the case for the massive quarks.
Namely, the fractional baryon number of the vacuum quark,
\begin{eqnarray}
   B_{q}(m, F) &=& - \frac{1}{2} \lim_{t \rightarrow 0} \sum_{n} \mbox{sgn}(E_{n}) e^{-t|E_{n}|}
 \label{eq : baryon_number0}  \\ \nonumber
                &=& \frac{1}{\pi} \left( F - \frac{1}{2}\sin 2F \right),
\end{eqnarray}
is independent of the quark mass in an analytical method by using the Debye expansion developed in \cite{Zahed}.
We also have checked that  this result is supported by the numerical computation by the Strutinsky method.
The detail procedure of the Debey expansion and the Strutinsky method for the massive quark will be discussed in the forthcoming paper.

Since the classical hedgehog solution is not an eigenstate of spin nor isospin, 
the states obtained above are not physical states.
In order to obtain physical states with proper quantum numbers, 
we need to quantize the classical hedgehog configuration 
through rotation in flavor $SU(3)$ space.
In this paper, however, we consider only the classical hedgehog solutions.
As we will see, the strength of the pion field is not very strong, where it is known that the effect of the rotation is not important for the mass of a quark droplet.

\subsection{Energy in the NJL chiral bag}

In this subsection, we discuss the energy of a quark droplet 
including the energies of quark vacuum, valence quarks 
and the pion cloud.  
In order to estimate the quark vacuum energy, we consider the 
following two components.  
One is the energy caused by the NJL interaction which is expressed by 
an effective bag constant $B$ and the other is the chiral Casimir energy.  

First, we evaluate an effective bag constant as the difference of the vacuum energies between the inside of the bag and the physical vacuum outside of the bag.
The effective bag constant is obtained by considering the sea quarks, which are interacting through the NJL interaction in the bag.
This is an empty bag in which sea quarks within the cutoff momentum contribute.
It is then measured from the vacuum energy of the bulk vacuum in the NJL model.
In the previous investigations \cite{Yasui, Yasui2}, it was shown that the effective bag constant in this procedure agreed with the phenomenological value which is used in the MIT bag model. 
The effective bag constant is given by 
\begin{eqnarray}
 B_{eff}(m,R) 
 &=& \sum_{q=u,d,s} \left[ \frac{(m_{q}-m_{0q})^{2}}{4G} - \frac{N_{c}}{V} \sum_{K^{P}} E_{K^{P}}(m_{q}, R) g(p_{q}/\Lambda) \right] - \epsilon_{0},
  \label{eq : effective_B}
\end{eqnarray}
where $V=(4\pi/3)R^{3}$ is the volume of the bag, $N_{c}$ the number of the degree of freedom of color.
Here $K^{P}$ refers to the eigenstate in the vacuum part in the bag for $q=u$, $d$ and $s$, respectively.
The vacuum quark contribution is cut off by a smooth regularization 
function of a Lorentzian type~\cite{Yasui2} with a momentum cutoff $\Lambda$
\begin{eqnarray}
 g(p/\Lambda) = \frac{1}{1+(p/\Lambda)^{a}}
  \label{eq : regular}
\end{eqnarray}
for smearing of the discrete energy levels of the quarks in the bag.
The smoothness parameter $a$ in the regularization (\ref{eq : regular}) is fixed to reproduce the nucleon mass $E_{N+\Delta}=1.1$ GeV.
We have obtained $a=22.58$.  
The last term is  the energy in the bulk vacuum $\epsilon_{0}$, 
in which chiral symmetry is maximally broken.
We mention that the effective bag constant is given at the zero chiral angle.
The effect of the pion cloud in the energy of the quark droplet is accounted 
by the chiral Casimir energy as discussed below.

The chiral Casimir energy 
has been studied extensively in 80's and the result is well known.
It is defined by
\begin{eqnarray}
  E_{C}(m, R, F) = - \frac{1}{2} \lim_{t \rightarrow 0_{+}} \sum_{n} \mbox{sign}(E_{n}) E_{n}  e^{-t|E_{n}|}.
  \label{defCenergy}
\end{eqnarray}
Here, the sum is taken over all the states with positive and negative energies.
In the massless case, it has been known that there is a logarithmic divergence in 
(\ref{defCenergy})~\cite{Vepstas, Zahed}.
In order to remove this divergence, we impose a condition that the second derivative of the chiral Casimir energy with respect to $F$ vanishes at $F=0$.
We found that the logarithmically divergent 
term is proportional to $\sin^{2}F$ for a finite quark mass by using the Debye expansion.
This analytical technique has been developed in the chiral bag for massless 
quarks~\cite{Zahed}.
Subtracting the divergent term, we find the subtracted finite Casimir energy,
\begin{eqnarray}
  E_{C}^{fin}(m, R, F) =  E_{C}(m, R, F) -  E_{C}(m, R, 0) - \frac{1}{2} \sin^{2}F \left. \frac{\partial^{2}E_{C}(m, R, F)}{\partial F^{2}} \right|_{F=0}.
    \label{eq : chiral_Casimir_energy3}
\end{eqnarray}
The reference point of the Casimir energy is set at $F=0$.
In order to calculate the sum numerically, the regularization, such as the exponential type, Gaussian type, heat kernel type \cite{Hosaka_Toki_86} and so on, have been used.
In our discussion, we use the Strutinsky smearing method \cite{Wust, Vepstas_Jackson}.
This method has an advantage that the maximum energy in the sum of 
(\ref{defCenergy}) to 
achieve good convergence can be taken relatively as small as $ER_{max} \sim 40$, 
while at least $ER_{max} \sim 100$ is necessary for other regulators.  
We note that $ER_{max} \sim 40$ is needed for the case of massive quarks, 
while even a smaller value $ER_{max} \sim 10$ is sufficient for massless quarks.
Also, we note that the chiral Casimir energy always takes 
a positive value for any quark mass and chiral angle.

Consequently, the total energy of the NJL chiral bag is given by a summation of the valence quark energy, the effective bag constant, the chiral Casimir energy and the pion cloud energy,
\begin{eqnarray}
   E_{tot}(m, R, F) = E_{val}(m, R, F) + B_{eff}(m, R)V + E_{C}^{fin}(m, R, F)+ E_{\pi}(R, F), 
   \label{eq : NJLCB}
\end{eqnarray}
where $m$ is the dynamical quark mass of either the $ud$ or $s$ quark. 
The first term indicates a contribution from the valence quarks.
We assume $N_{u}=N_{d}$ in the discussion below, and further ignore 
the Coulomb energy, since we consider only small $A$ system.  

The input parameters in the calculation of the energy 
of a quark droplet are the baryon number $A$ and 
the strangeness $S$.
For a given baryon number $A$ and a strangeness $S$, 
$N_{u}+N_{d} = N_{c}A+S$ and $N_{s}=-S$.  
The energy is then obtained by taking a variation 
with respect to the dynamical quark mass 
$m_{u}=m_{d}$ and $m_{s}$, the chiral angle $F$ at the bag surface, 
and the bag radius $R$.

\section{Numerical Result}

\subsection{Quark energy level in the NJL chiral bag}

\begin{figure}[tbp]
\begin{center}
\includegraphics[width=14cm, keepaspectratio, clip]{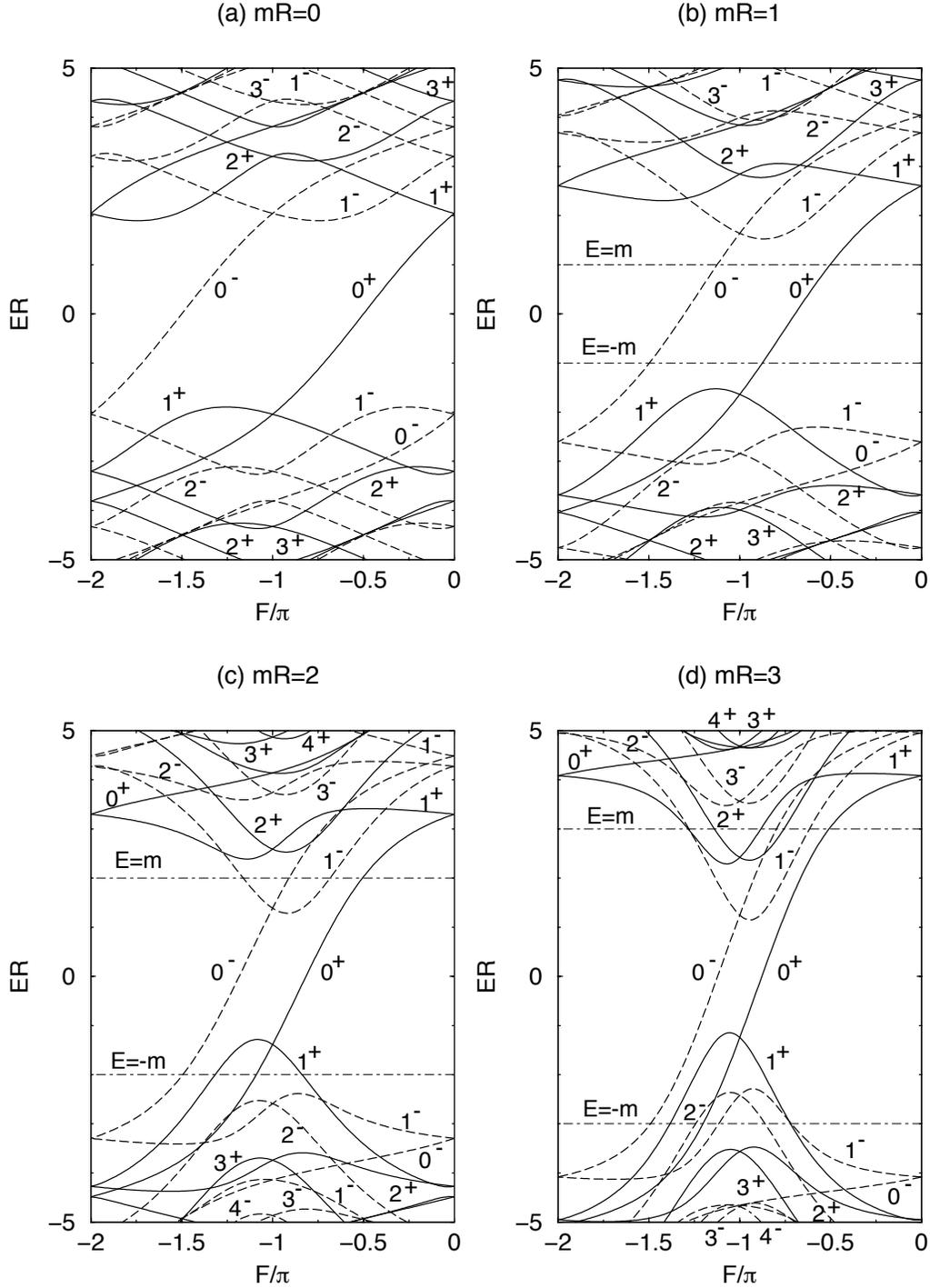}
\end{center}
\vspace*{-1.0cm} \caption{\small \baselineskip=0.5cm The energies $E$ of various quark levels in a chiral bag as functions of the chiral angle  $F$. (a) is for $mR=0$, (b) $mR=1$, (c) $mR=2$ and (d) $mR=3$, where $m$ is the $ud$ quark mass and $R$ the bag radius. The horizontal dot-dashed lines indicate $|E|=m$.
The $K$-grand spins are denoted for $K=0, 1, 2, 3$ and $4$.
The $\pm$ parity is distinguished by the solid ($P=+1$) and dashed ($P=-1$) lines.}
 \label{fig : ER_F_mR0_KP}
\end{figure}

The energy levels of $ud$ quarks in the chiral bag 
are affected by the chiral angle at the bag surface $F$.
By solving the boundary condition Eq.~(\ref{eq : boundary_vector_2}), 
we obtain the energy levels which are shown as functions of $F$ in Fig.~\ref{fig : ER_F_mR0_KP}.
The four figures correspond to the quark mass $mR=0$, $1$, $2$ and $3$, respectively.
The $K$-grand spins are denoted for $K=0, 1, 2, 3$ and $4$.
The solid and dashed lines are for the positive 
and negative parity states.
Horizontal dot-dashed lines are for $|E|=m$.

It is one of the particular points in the chiral bag model that the first $0^{+}$ states dives into the vacuum  
as the absolute value of the chiral angle $F$ increases.
The critical chiral angle for $E=0$ in the $0^{+}$ 
state is $F/\pi=-0.5$, $-0.70$, $-0.82$ 
and $-0.87$ for $mR=0$, $1$, $2$ and $3$, respectively.
As the quark mass increases, a larger chiral angle is 
necessary for pulling the valence quark down into the vacuum part.
For $0^{+}$ state, we obtain $E=m$ at $F/\pi=-0.5$ for any quark mass.
The states in $|E|<m$ are obtained by substituting the momentum 
with an imaginary value $ip$ in Eq.~(\ref{eq : boundary_vector_2}).
As the quark mass increases, more states appear in the energy region of 
$|E|<m$, or $|F| \sim \pi$.  
For example, we have only the lowest $0^{\pm}$ states in this region for $mR=1$, 
while $0^{\pm}$ and $1^{\pm}$ states are also included in $|E|<m$ for $mR=2$.
For $mR=3$, $2^{\pm}$ states are also included in $|E|<m$.

In the massless fermion, there is a symmetry of the energy level 
for $F \rightarrow F + \pi$ under the change in the parity.
However, the massive fermion does not have this symmetry.
In order to see the origin of this asymmetry, 
we write the equation of motion for the quark as
\begin{eqnarray}
\left[  
i \partial\hspace{-0.2cm}/ - m_{h} 
-\frac{1}{2}e^{i\vec{\tau}\cdot\vec{n}\gamma_{5}F(R)} \delta(r-R)
 \right]h = 0,
\end{eqnarray}
where the boundary condition is included as the delta function term.
The sign of the $\delta$-function term
changes from negative to positive as the chiral angle changes 
from $F=0$ to $-\pi$.
Therefore, the sum of the mass term and the scalar term proportional to $\cos F$ in the $\delta$-function changes 
from 
$m_{h} + \delta(r-R)/2$ to $m_{h} - \delta(r-R)/2$ for $F=0$ 
to  $-\pi$.
This is the origin of the asymmetry of the quark levels.
These quantities can be interpreted 
as an effective mass, or the scalar potential, of the quark in the chiral bag.

\subsection{Multi-quark droplets}

Now, we discuss the stability of quark droplets.
We are interested in the quark droplets 
with small baryon numbers up to $A \simeq 5$, 
in which the surface pion is expected to have a non-negligible contribution.
To start with, we discuss the procedure of energy variation of a quark droplet.
We have several dynamical variables 
in the energy (\ref{eq : NJLCB}) of a quark droplet 
with baryon number $A$ and strangeness $S$; 
the dynamical quark mass ($m_{u}=m_{d}$ and $m_{s}$), 
the chiral angle $F$ at the bag surface and the bag radius $R$.
As shown in our previous work in the NJL model with the MIT bag \cite{Yasui2},  variation of the total energy with respect to 
the dynamical quark mass leads to 
the restoration of chiral symmetry 
for $R \ltap 6$ fm for $ud$ quark sector ($m_{ud} = 0$), 
and $R \ltap 3$ fm for the $s$ quark sector ($m_{s} = m_{0s}$).
This is also the case in the NJL model with the chiral bag.

\subsubsection*{$A = 1$}

First, we discuss a strangelet of the baryon number $A=1$, namely the nucleon.
As mentioned above, since the chiral condensate $\langle \bar{q} q \rangle$ vanishes, the results of the present model 
differs from the conventional bag model only by the bag constant 
term.  
The stable solution for a given bag radius $R$ is then obtained 
by taking a variation of the total energy 
with respect to $F$.
We show the numerical result in Fig.~\ref{fig : E_R_B}(a), where
the total energy as well as 
various contributions, such as the pion cloud energy, 
the quark energy (the sum of the valence and Casimir energies), 
the Casimir energy and the volume energy of the effective bag constant, 
are shown as functions of $R$.
The total energy has two local minima 
at $R=0$ and $0.750$ fm.
The first local minimum corresponds to the pure Skyrmion, 
in which the quark contribution vanishes.
The second minimum is characteristic in the present model, where both the pion and quarks contribute to the physical quantities.
The chiral angle is shown as a function of the bag radius 
by the solid line in Fig.~\ref{fig : F_R}.
At $R=0.750$ fm, we obtain $F/\pi=-0.327$.

The appearance of the second local minimum is due to the $R$ dependence of the effective bag constant $B_{eff}$ \cite{Yasui2}.
Here, we note that the minimum value of the volume energy $B_{eff}V$ depends on the smoothness parameter $a$ in the regularization, while the corresponding bag radius does little.
We have used that dependence in order to reproduce the mass of the nucleon.
At this point, the resulting volume energy takes either positive or negative value.
(In the present model, it takes a negative value.)
However, its absolute value is not large, and the most part of the mass of the nucleon is given by the quark and the pion energies; $E_{\pi} = 0.437$ GeV and $E_{q}=0.715$ GeV.

\subsubsection*{$A = 2$}

Next, let us discuss the case of baryon number $A=2$.
We consider several configurations for $ud$ quarks for a fixed strangeness $S$.
The $ud$ quarks can occupy various levels associated with the chiral angle as shown in Fig~\ref{fig : ER_F_mR0_KP}(a).
The first three quarks occupy the $0^{+}$ state, and form a closed shell.
For the remaining three quarks, the $1^{+}$ state is available for $F/\pi > -0.317$, while the $1^{-}$ state is also used for $-1\le F/\pi < -0.317$.
Therefore, we shall compare the following two configurations.

For $S=0$, we consider a) $(0^{+})^{3}(1^{+})^{3}$ and b) $(0^{+})^{3} \, (1^{-})^{3}$.
Here $(K^{P})^{N}$ denotes a configuration how $N$ quarks accommodate the state $K^P$.  
The state $K^{P}$ can contain $N_{c} \, (2K+1)$ $ud$ quarks maximally.
By comparing the minimum energy of the configurations a) and b), we find that the configuration a) is more stable.
The reason that we consider two configurations a) and b) is that lower quark levels depend on the chiral angle.
We also note that the parity affects for a) and b).
Here, we simply find a configuration which has the minimum energy.

\begin{figure}[tbp]
\begin{minipage}{8cm}
\vspace*{0.0cm}
\centering
\includegraphics[width=8cm]{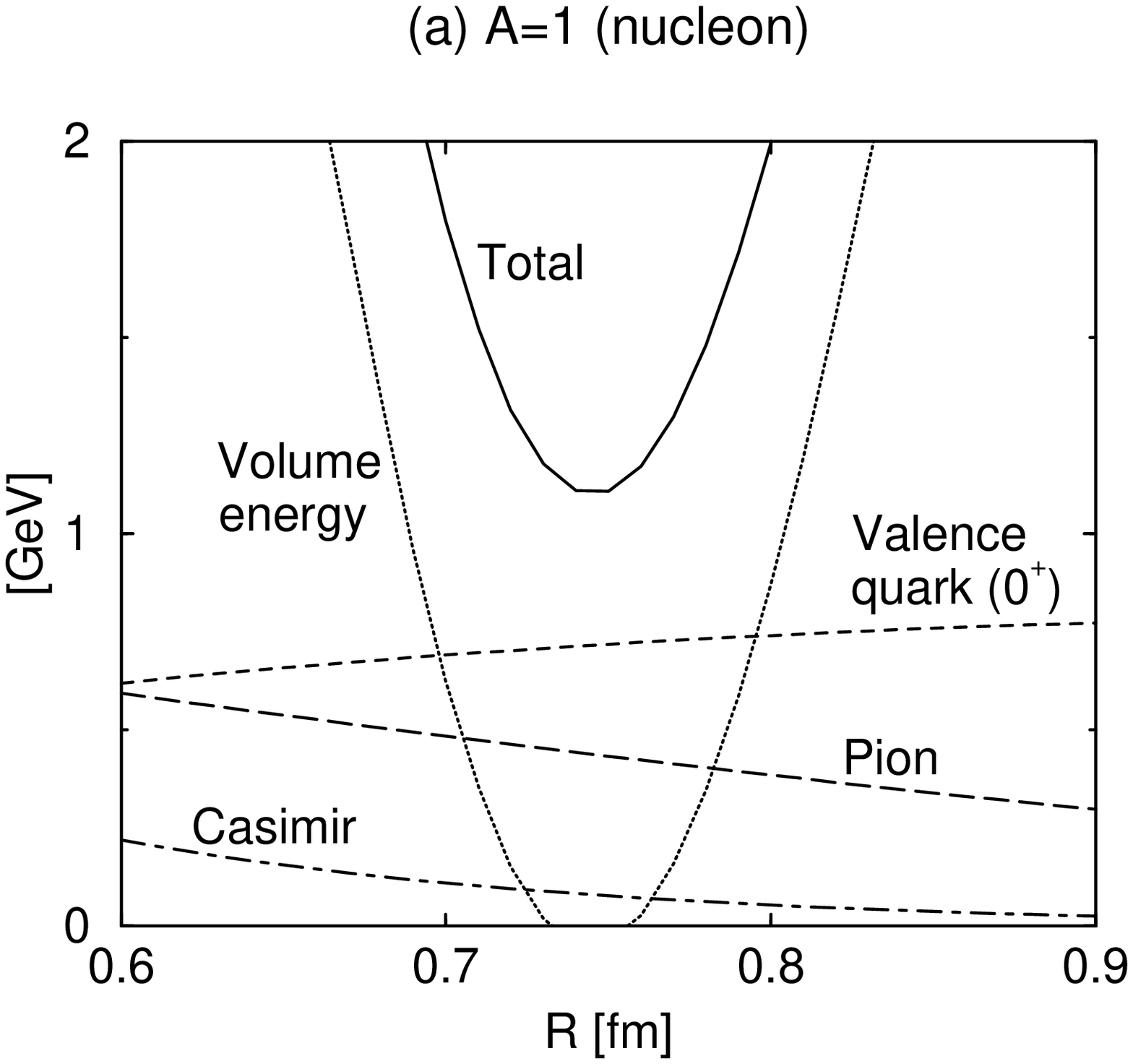}
\vspace{-0.0cm}
\end{minipage}
\begin{minipage}{8cm}
\centering
\includegraphics[width=8cm]{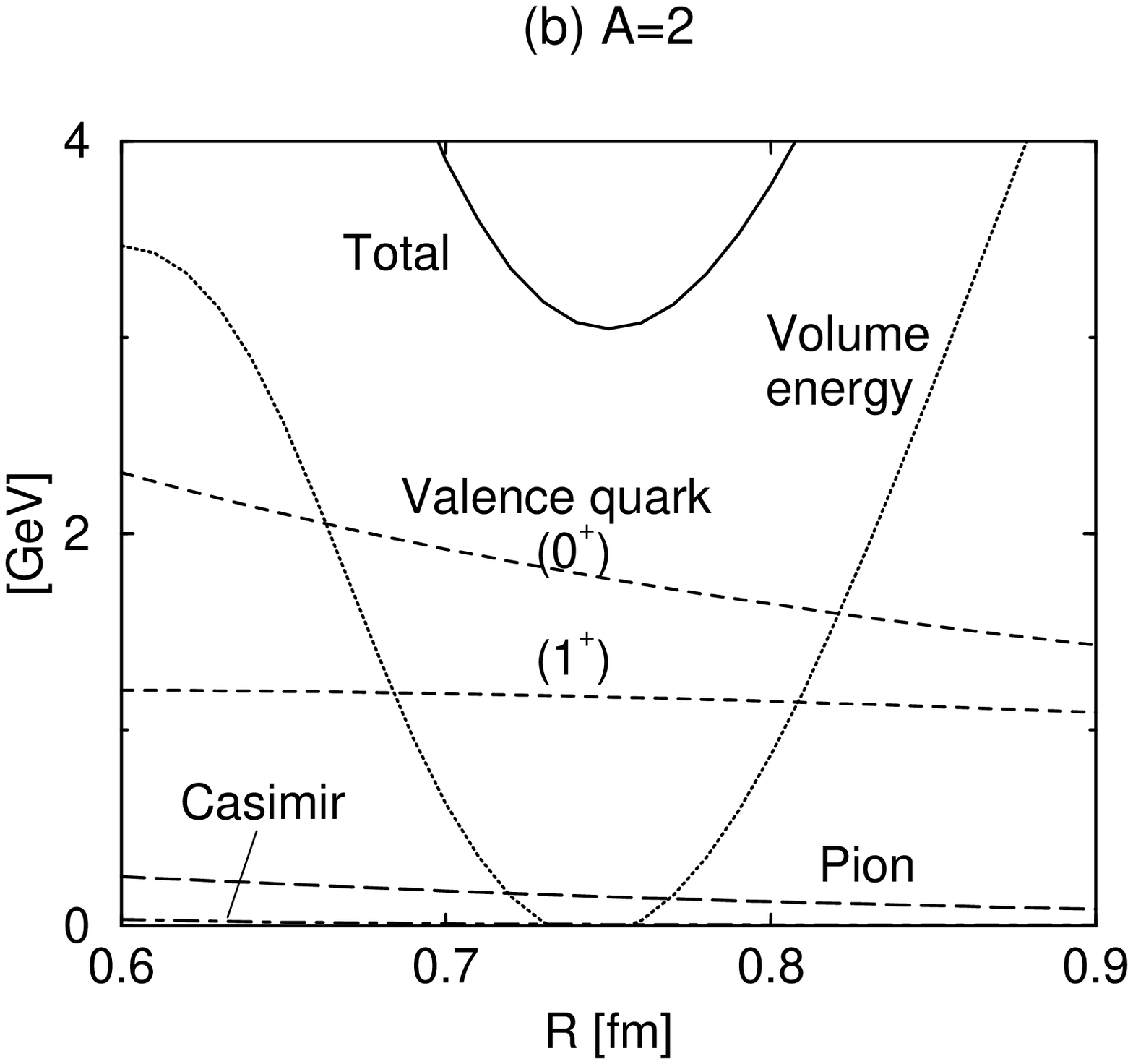}
\vspace{-0cm}
\end{minipage}
\caption{\small \baselineskip=0.5cm Various energy contributions to a quark droplet of baryon number $A=1$ (nucleon) and $A=2$.}
    \label{fig : E_R_B}
\end{figure}

\begin{figure}[tbp]
\begin{center}
\includegraphics[width=8cm, keepaspectratio, clip]{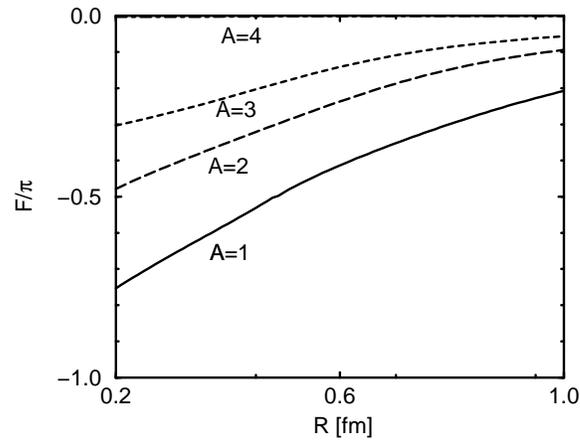}
\end{center}
\vspace*{-1.0cm} \caption{\small \baselineskip=0.5cm The chiral angle $F$ at the bag surface as functions of the bag radius $R$ for non-strange quark droplets of the baryon number $A=1$, 2, 3 and 4.}
 \label{fig : F_R}
\end{figure}

In Fig.~\ref{fig : E_R_B}(b), we show the total energy as well as various components such as the quark energy (the valence quark plus the chiral Casimir energy), the Casimir energy, the pion cloud and the volume energy for $ud$ quark configuration $(0^{+})^{3}(1^{+})^{3}$ .
We find a local minimum with the energy $E=3.042$ GeV at the bag radius $R=0.750$ fm, where 
the total energy is dominated by the valence quarks.
They are $E_{3h, 0^{+}}=1.160$ GeV and $E_{3h, 1^{+}}=1.770$ GeV for the $0^{+}$ and $1^{+}$ states occupied by three quarks, respectively, while the pion cloud energy is $E_{\pi}=0.148$ GeV which is smaller than that in the nucleon ($A=1$).
The small pion contribution is due to the small chiral angle in absolute value.
As shown in Fig.~\ref{fig : F_R}, we find $F/\pi=-0.165$ at the stable point, which is smaller than that of the nucleon $F/\pi=-0.327$.

For a finite strangeness, we investigate the configurations a) $(0^{+})^{3} \, (1^{+})^{3-|S|}$ and b) $(0^{+})^{3} \, (1^{-})^{3-|S|}$ for $S=-1$ and $-2$, respectively.
Again, the configuration a) is found to be more stable for each strangeness.
We summarize the numerical results in Table \ref{tbl : A2}, where
we show the $ud$ quark configurations, the energy per baryon number $E/A$, the bag radius $R$, the chiral angle $F$ and the pion energy $E_{\pi}$ for each strangeness.
The energy per baryon number is $E/A=1.521$, $1.527$ and $1.526$ GeV for $S=0$, $-1$ and $-2$, respectively.
Therefore, the most stable state is the non-strange state of $S=0$, when 
the bag radius is $R=0.750$ fm, and the chiral angle $F/\pi=-0.165$. 

To see the reason that the non-strange state is more stable, let us look at the energy levels of Fig.~1(a).
At the chiral angle $F/\pi=-0.165$ where $R=0.750$ fm, 
among the six $ud$ quarks, the three $ud$ quarks occupy the $0^{+}$ state, and the other three occupy the $1^{+}$ state with $E_{h, 1^{+}}=2.24/R$.
The latter should be compared with the energy of the strange quark of $1/2^{+}$ state, $E_{s, 1/2^{+}}=2.30/R$.
Hence $s$ quark state is above the $ud$ quark $1^{+}$ state, and
therefore, the strangeness cannot be included in the ground state.

From the above discussions, we have seen that the energy per baryon number $E/A=1.521$ GeV of the quark droplet is larger than the nucleon mass $E_{N+\Delta}=1.1$ GeV.
Therefore, our result indicates that the quark droplet with $A=2$ is not stable against the decay into two nucleons.

\subsubsection*{$A = 3$ {\rm and} $4$}

For $A=3$, we investigate the following $ud$ quark configurations, a) $(0^{+})^{3} \, (1^{+})^{6-|S|}$ and b) $(0^{+})^{3} \, (1^{-})^{6-|S|}$ for $S=0$, $-1$, $-2$ and $-3$, and found  
that the configuration a) is more stable than b) for each strangeness.
We show the numerical results in Table \ref{tbl : A3}.
We obtained the energy per baryon number $E/A=1.588$, $1.605$, $1.618$ and $1.630$ GeV for $S=0$, $-1$, $-2$ and $-3$, respectively.
Therefore, the most stable state is the non-strangeness state with $S=0$.
There, the bag radius is $R=0.754$ fm, and the chiral angle is $F/\pi=-0.0580$.
The chiral angle is very small as compared with that of the nucleon.
Correspondingly, the pion cloud energy is $E_{\pi}=0.02267$ GeV, which is very small as compared with that of the nucleon.

In order to understand the reason that the strange quark is not contained in the ground state, we turn to the $ud$ quark energy level in Fig.~\ref{fig : ER_F_mR0_KP}(a).
At the chiral angle $F/\pi=-0.0580$, the three $ud$ quarks occupy the lowest $0^{+}$ state.
The other six $ud$ quarks occupy the $1^{+}$ state with eigenenergy $E_{1^{+}}=2.24/R$, which is lower than the lowest $s$ quark energy $E_{s,1/2^{+}}=2.30/R$.
The $1^{-}$ energy is also lower than the $s$ quark energy, and therefore, 
the $s$ quark does not appear in the ground state of the quark droplet for $A=3$.

For the baryon number $A=4$, we considered the configurations a) $(0^{+})^{3} \, (1^{+})^{9-|S|}$ and b) $(0^{+})^{3} \, (1^{-})^{9-|S|}$ for the strangeness $S=0$, $-1$, $-2$, $-3$ and $-4$.
We have found again that the configuration a) is more stable for each strangeness.
We show the numerical result in Table \ref{tbl : A4}, where we see that 
the energies per baryon number are $E/A=1.597$, $1.616$, $1.630$, $1.644$ and $1.731$ GeV for $S=0$, $-1$, $-2$, $-3$ and $-4$, respectively.
Hence, the non-strange state of $S=0$ is the most stable one, when the bag radius is $R=0.757$ fm and the chiral angle is $F/\pi=-0.144 \times 10^{-3}$.
Here, the chiral angle is negligibly small, and so is the pion cloud energy.
In this way, it is justified to use the MIT bag model instead of the chiral bag model for $A=4$ and larger \cite{Yasui, Yasui2} (see also the discussion below).

\subsubsection*{$A = 5$}

Lastly, we consider the case of baryon number $A=5$.
We consider the $ud$ quark configurations a) $(0^{+})^{3} \, (1^{+})^{9} \, (1^{-})^{3-|S|}$ and b) $(0^{+})^{3} \, (1^{-})^{9} \, (1^{+})^{3-|S|}$ and c) $(0^{+})^{3} \, (1^{-})^{9} \, (0^{-})^{3-|S|}$ for $S=0, -1, -2$, and a) $(0^{+})^{3} \, 
(1^{+})^{9-|S|}$ and b) $(0^{+})^{3} \, (1^{-})^{9-|S|}$ for $S=-3, -4, -5$.
Among them, the configuration a) is the most stable one for each strangeness.
In Table \ref{tbl : A5}, we show the numerical results.
The energy of the quark droplet is $E/A=1.762$, $1.725$, $1.683$, $1.638$, $1.710$ and $2.226$ GeV for $S=0, \cdots, -5$, respectively.
Therefore, the most stable state is $S=-3$ of finite strangeness at the bag radius $R=0.761$ fm, where the chiral angle and the pion cloud energy are almost zero.

We can clarify the reason why the finite strangeness is included in the ground state of the quark droplet with $A=5$.
Let us turn again to the $ud$ quark energy level in Fig. \ref{fig : ER_F_mR0_KP}(a).
At the chiral angle $F \simeq 0$, the three quarks occupy the $0^{+}$ state with eigenenergy $E_{h, 0^{+}} \simeq 2.04/R$, and the nine quarks form a closed shell in the $1^{+}$ state with $E_{h, 1^{+}} \simeq 2.04/R$.
The remaining three quarks may occupy the $1^{-}$ state with $E_{h, 1^{-}} \simeq 3.20/R$, which is larger than the strange quark energy $E_{s, 1/2^{+}}=2.31/R$.
Therefore, the three quarks occupy the $1/2^{+}$ state in the $s$ quark sector.
Concerning the chiral angle $F$, it is exactly the same for the case of $A=4$, since the $ud$ quark configurations are the same.
Therefore, it is a good approximation to neglect the pion cloud for multi-baryon system.

For the baryon number $A \gtap 5$, the results of the NJL chiral bag model becomes similar to the NJL model with the MIT bag \cite{Yasui, Yasui2}.
As discussed in our previous work \cite{Yasui2}, the Multiple Reflection Expansion (MRE) reproduces the energy per baryon number obtained by solving the discrete quark levels for the baryon number $A \gtap 10$.
There, it was concluded that the strangelets with baryon number $10 \ltap A \ltap 10^{3}$ were not stable as compared with the normal nuclei \cite{Yasui}.
In this paper, we have verified again that the NJL chiral bag model does not support a stable strangelet as the ground state of the finite quark system.
This conclusion as derived in our previous works is qualitatively different from the analysis by the MIT bag model \cite{Witten, Farhi, Madsen}.

\begin{table}[htdp]
\begin{center}
\begin{tabular}{|c|l|c|c|c|c|} \hline
 $S$ &  $(K^{P})^{\#(\mbox{\small $ud$ quarks})}$ & $E$ ($E/A$) [GeV]  &  $R$ [fm]  & $F/\pi$ & $E_{\pi}$ [GeV] \\  \hline
   0   &  \hspace{0.5cm} $(0^{+})^{3} \, (1^{-})^{3}$ & 3.042 (1.521) & 0.750 &  -0.165  &  0.1480  \\ \hline
  -1   &  \hspace{0.5cm} $(0^{+})^{3} \, (1^{-})^{2}$ & 3.054 (1.527) & 0.750 &  -0.197  &  0.2117  \\ \hline
  -2   &  \hspace{0.5cm} $(0^{+})^{3} \, (1^{-})^{1}$ & 3.052 (1.526) & 0.749 &  -0.230  & 0.2856  \\ \hline
\end{tabular}
\end{center}
\caption{\small The $ud$ quark configurations, the total energy $E$, the bag radius $R$, the chiral angle $F$ and the pion cloud energy $E_{\pi}$ for quark droplets $A=2$ with several strangeness $S$. The values in the parentheses in the third row are the energies per baryon number $E/A$.}
\label{tbl : A2}
\end{table}%

\begin{table}[htdp]
\begin{center}
\begin{tabular}{|c|l|c|c|c|c|} \hline
 $S$ &  $(K^{P})^{\#(\mbox{\small $ud$ quarks})}$ & $E$ ($E/A$) [GeV]  &  $R$ [fm]  & $F/\pi$ & $E_{\pi}$ [GeV] \\  \hline
   0   &  \hspace{0.3cm} $(0^{+})^{3} \, (1^{+})^{6}$ & 4.764 (1.588) & 0.754 &  -0.0580  &  0.02267  \\ \hline
  -1   &  \hspace{0.3cm} $(0^{+})^{3} \, (1^{+})^{5}$ & 4.815 (1.605) & 0.754 &  -0.0811 &  0.04429  \\ \hline
  -2   & \hspace{0.3cm}  $(0^{+})^{3} \, (1^{+})^{4}$ & 4.854 (1.618) & 0.754 &  -0.105  & 0.07527  \\ \hline
  -3   & \hspace{0.3cm}  $(0^{+})^{3} \, (1^{+})^{3}$ & 4.890 (1.630) & 0.753 &  -0.131  & 0.1169  \\ \hline
\end{tabular}
\end{center}
\caption{\small Same as Table 1 for $A=3.$}
\label{tbl : A3}
\end{table}%

\begin{table}[htdp]
\begin{center}
\begin{tabular}{|c|l|c|c|c|c|} \hline
 $S$ &  $(K^{P})^{\#(\mbox{\small $ud$ quarks})}$ & $E$ ($E/A$) [GeV]  &  $R$ [fm]  & $F/\pi$ & $E_{\pi}$ [GeV] \\  \hline
   0   &  \hspace{0.3cm} $(0^{+})^{3} \, (1^{+})^{9}$ & 6.388 (1.597) & 0.757 &  $-0.144 \times 10^{-3}$  &  $0.1675 \times 10^{-6}$  \\ \hline
  -1   &  \hspace{0.3cm} $(0^{+})^{3} \, (1^{+})^{8}$ & 6.464 (1.616) & 0.750 &  -0.0151 &  0.001828  \\ \hline
  -2   & \hspace{0.3cm}  $(0^{+})^{3} \, (1^{+})^{7}$ & 6.520 (1.630) & 0.757 &  -0.0308  & 0.007702  \\ \hline
  -3   & \hspace{0.3cm}  $(0^{+})^{3} \, (1^{+})^{6}$ & 6.576 (1.644) & 0.757 &  -0.0481  & 0.01876  \\ \hline
  -4   & \hspace{0.3cm}  $(0^{+})^{3} \, (1^{+})^{5}$ & 6.924 (1.731) & 0.758 &  -0.0667  & 0.03612 \\ \hline
\end{tabular}
\end{center}
\caption{\small Same as Table 1 for $A=4$.}
\label{tbl : A4}
\end{table}%

\begin{table}[htdp]
\begin{center}
\begin{tabular}{|c|l|c|c|c|c|} \hline
 $S$ &  $(K^{P})^{\#(\mbox{\small $ud$ quarks})}$ & $E$ ($E/A$) [GeV]  &  $R$ [fm]  & $F/\pi$ & $E_{\pi}$ [GeV] \\  \hline
   0   & \hspace{0.2cm}   $(0^{+})^{3} \, (1^{+})^{9} \, (1^{-})^{3} $ & 8.630 (1.762) & 0.762 &  -0.119  &  0.09780  \\ \hline
  -1   & \hspace{0.2cm}  $(0^{+})^{3} \, (1^{+})^{9} \, (1^{-})^{2} $  & 8.625 (1.725) & 0.762 &  -0.0702  &  0.03348  \\ \hline
  -2   & \hspace{0.2cm}  $(0^{+})^{3} \, (1^{+})^{9} \, (1^{-})^{1} $  & 8.415 (1.683) & 0.761 &  -0.0200  &  0.03820  \\ \hline
  -3   & \hspace{0.2cm}  $(0^{+})^{3} \, (1^{+})^{9}                      $  & 8.190 (1.638) & 0.761 &  $-0.854 \times 10^{-4}$  &  $0.6923 \times 10^{-7}$  \\ \hline
  -4   & \hspace{0.2cm}  $(0^{+})^{3} \, (1^{+})^{8}                      $  & 8.550 (1.710) & 0.761 &  -0.0127  &  0.001536  \\ \hline
  -5   & \hspace{0.2cm}  $(0^{+})^{3} \, (1^{+})^{7}                      $  & 11.13 (2.226) & 0.762 &  -0.0304  & 0.007548  \\ \hline
\end{tabular}
\end{center}
\caption{\small Same as Table 1 for $A=5$.}
\label{tbl : A5}
\end{table}%

\section{Conclusion}

In this paper, we discussed the stability of strangelets by considering dynamical chiral symmetry breaking of QCD.
We investigated the effects of the dynamical generation of quark masses in a finite volume by introducing the NJL model inside the chiral bag.  
This is a chirally symmetric two phase model, in which quarks inside the bag are interacting through the point-like interaction of the NJL model, and mesons exist outside the bag as meson cloud.

We adopted the hedgehog configuration to calculate the energies of the systems of  baryon number $A \leq 5$.  
The dynamical quark masses of the $u$, $d$ and $s$ quarks were then given in the mean field approximation of the NJL interaction.
It turned out that for strangelets with small baryon number, chiral symmetry is restored inside the bag, and hence the dynamical mass of the quarks vanishes.  
However, for general purposes, we have investigated various chiral Casimir effects with finite quark mass.  
By using analytical and numerical methods, it was verified that the fractional baryon number in the $ud$ quark sector is not affected by the finite quark mass.
Therefore, the total baryon number is conserved exactly, as in the case of massless quarks.  
The Casimir energy was also well defined by subtracting the divergent term for the massive quark.

The total energy of a quark droplet was expressed as a sum of the energy of the valence quarks, the Casimir energy, the volume energy of the effective bag constant and the pion cloud energy.
The energy of the system was measured from the reference point of the chirally broken vacuum outside the bag.
The energy of the quark droplet with baryon number $A$ was determined by taking a variation with respect to (in addition to the dynamical quark masses) the bag radius, the chiral angle and the strangeness.
As a result, we could reproduce the nucleon mass $E=1.1$ GeV and the nucleon radius $R=0.750$ fm.
For the multi-baryon states, it was shown that the quark droplets with baryon number $A=2$, $3$ and $4$ did not contain the strangeness in the ground state.
For the baryon number $A \ge 5$, the quark droplets contained the finite strangeness.
We obtained the strangeness $S=-3$ for $A=5$.  
For complication of analysis of discrete states, we did not investigated explicitly systems of $A > 5$, but we expect that for larger $A$ strangelets of finite strangeness will become the ground states.  

We introduced the Lorentzian regularization for the quark vacuum energy associated with the NJL interaction.
This quantity was  expressed by an effective bag constant.  
We have also investigated the Gaussian type regularization instead of the Lorentzian.
We found that the Gaussian regularization gave the effective bag constant which was larger than the value used in the conventional MIT bag model, which is consistent with our previous result~\cite{Yasui2}.
Accordingly, the mass of the nucleon became too large and could not reproduce the observed value.  

In the present study, we have found that the pion contribution was suppressed as the baryon number increased.
The pion cloud energy in the quark droplet of $A=5$ (and likely for larger $A$) was almost negligible.
In this way, the use the MIT bag model for the quark droplets with baryon number $A \gtap 5$ \cite{Yasui, Yasui2} can be justified.
The role of the pion cloud was previously investigated for non-strange multi-baryon systems with $A=2$, $3$ and $4$ in chiral bag model \cite{Vento, Wuest, Takashita, Blaettel}.
Our result is consistent with theirs.

The energies per baryon of these quark droplets are larger than the mass of the nucleon.
Therefore, the strangelets are not stable against decays into multi-baryons.
This result is qualitatively different from the previous studies in the MIT bag model.

There are several issues to be further discussed.
In this paper, we have assumed the hedgehog ansatz at the classical level.
For realistic comparison with experimental data, it is necessary to perform spin-isospin projection.  
It is also interesting to discuss the Nambu-Goldstone mode in the chiral symmetry broken phase in the bag for large $A$ system.  
There, the Nambu-Goldstone mode will directly couple to the quarks inside the bag,  not only at the bag surface.
The vacuum structure in the bag may be affected by this volume type coupling.
We are now currently working on this problem.
The explicit $U(1)_{A}$ breaking is also an important topic.
There, the 't Hooft term will be introduced in addition to (or instead of) the NJL interaction.
The effect of the $qq$ correlation is also an interesting subject \cite{Madsen01, Amore, Kiriyama}.
As the baryon number increases, the number of the valence quarks also increases.
There, it would be possible to cause the color superconductivity.
It is an interesting question whether the $qq$ correlation in the bag is affected by the pion cloud outside the bag.

\section*{Acknowledgement}
We acknowledge valuable discussions with Prof. H.~Toki.
We are very grateful to Prof. T.~Kunihiro for discussions and encouragements.
We also express our thanks to Prof. M.~Oka for useful discussions.
We would like to thank Dr. Y.~Ogawa and Dr. N.~Ikezi for comments.
Part of this study was proceeded when S. Y.  was belonging to
Yukawa Institute for Theoretical Physics, Kyoto University.
This work was also partially supported by Grant-in-Aid for Scientific Research for Priority Areas, MEXT (Ministry of Education, Culture, Sports, Science and Technology) with No. 17070002.


\end{document}